\documentclass[aps, a4paper,superscriptaddress,12pt,preprintnumbers,floatfix,nofootinbib]{revtex4-1}
\usepackage{url}
\usepackage{xcolor}
\usepackage{hyperref}
\usepackage[normalem]{ulem}
\hypersetup{colorlinks=true,linkcolor=redLinks,citecolor=greenLinks,
urlcolor=redLinks,
pdfborder={0 0 1}}
\hypersetup{
    bookmarks=true,         
    unicode=false,          
    pdftoolbar=true,        
    pdfmenubar=true,        
    pdffitwindow=false,     
    pdfstartview={FitH},    
    pdftitle={My title},    
    pdfauthor={Author},     
    pdfsubject={Subject},   
    pdfcreator={Creator},   
    pdfproducer={Producer}, 
    pdfkeywords={keyword1, key2, key3}, 
    pdfnewwindow=true,      
    colorlinks=false,       
    linkcolor=red,          
    citecolor=green,        
    filecolor=magenta,      
    urlcolor=cyan           
}
\colorlet{shadecolor}{gray!15}
\definecolor{greenLinks}{rgb}{0, 0.6, 0} 
\definecolor{blueLinks}{rgb}{0, 0, 0.6}
\definecolor{redLinks}{rgb}{0.6, 0, 0}
\definecolor{tempText}{rgb}{0.55, 0.10,0.67}
\definecolor{eprintLinks}{rgb}{0.4, 0.4, 0.4}
\definecolor{journalLinks}{rgb}{0.6, 0, 0}
\newcommand{\MYhref}[3][redLinks]{\href{#2}{\color{#1}{#3}}}%
\usepackage{subfigure}
\usepackage{rotating}
\usepackage{slashed}     

\usepackage[T1]{fontenc} 
\usepackage[utf8]{inputenc}
\usepackage[english]{babel}
\usepackage{amsmath}
\usepackage{graphicx}
\usepackage{color}
\usepackage{mathrsfs}   
\usepackage{amssymb}
\usepackage{hyperref}
\usepackage{dcolumn}
\usepackage{bm}
\usepackage{graphicx}

\def\lnv{lepton number violation }
\def\vev#1{\left\langle #1\right\rangle}
\newcommand{\sm}{Standard Model }
\def\SM{$\mathrm{SU(3)_c \otimes SU(2)_L \otimes U(1)_Y}$ }
 
\def\TrTrOne{ $\mathrm{SU(3)_c \otimes SU(3)_L \otimes U(1)_X}$ }
\newcommand{\AddrAHEP}{
  {\it AHEP Group, Instituto de F\'{\i}sica Corpuscular --
    C.S.I.C./Universitat de Val{\`e}ncia \\
    Edificio de Institutos de Paterna,
 C/Catedratico Jos\'e Beltr\'an, 2 \\E-46980 Paterna (Val\`{e}ncia) - SPAIN}}

\begin{document}

\title{CP violation from flavor symmetry \\
in a lepton quarticity dark matter model  } 

\author{Salvador Centelles Chuli\'a}
\email{salcen@alumni.uv.es} \affiliation{\AddrAHEP} \author{Rahul
  Srivastava} \email{rahuls@prl.res.in}
\affiliation{ Physical Research Laboratory, Navrangpura, Ahmedabad - 380 009, INDIA \\ \& \\
  The Institute of Mathematical Sciences, Chennai 600 113, INDIA }
\author{Jos\'e W.F. Valle} \email{valle@ific.uv.es}
\affiliation{\AddrAHEP} \vspace{1cm} 
\begin{abstract}
   \vspace{1cm}
   
   We propose a simple $\Delta (27) \otimes Z_4$ model where neutrinos
   are predicted to be Dirac fermions. The smallness of their masses
   follows from a type-I seesaw mechanism and the leptonic CP
   violating phase correlates with the pattern of $\Delta (27)$ flavor
   symmetry breaking.  The scheme naturally harbors a WIMP dark matter
   candidate associated to the Dirac nature of neutrinos, in that the
   same $Z_4$ lepton number symmetry also ensures dark matter
   stability.

\end{abstract}

\maketitle


\section{Introduction}
\label{sec:introduction}


Currently the information on neutrino properties comes mainly from
neutrino oscillation experiments~\cite{Maltoni:2004ei}.
These are insensitive to whether neutrinos are Dirac or Majorana
fermions~\cite{Schechter:1980gr,Schechter:1980gk,Doi:1980ze}.
The fact that the weak interaction is V-A turns the quest for \lnv
and the Majorana nature of neutrinos into a major experimental
challenge~\cite{avignone:2007fu,Barabash:2004pu,Blot:2016cei}.
The detection of neutrinoless double beta decay would signify a major
step forward in this endeavor. According to the
  black--box theorem~\cite{Schechter:1981bd,Duerr:2011zd} its
  observation would demonstrate that neutrinos are Majorana fermions
  and thus lepton number is violated in nature.

Concerning the mechanism responsible for generating small neutrino
masses, little is known regarding the nature of its associated
messenger particles, their characteristic mass scale or other detailed
features of the effective operator~\cite{Valle:2015pba}.
The smallness of neutrino masses almost always assumes them to be
Majorana fermions. For example, this is the case in the conventional
high--scale
(type-I)~\cite{gell-mann:1980vs,yanagida:1979,mohapatra:1980ia,Schechter:1980gr}
seesaw mechanism.
Likewise in low--scale variants of the seesaw mechanism as well as in
radiative schemes, neutrinos turn out to be Majorana fermions, as
reviewed in~\cite{Boucenna:2014zba}.

Having naturally light Dirac neutrinos requires extra assumptions
beyond the standard \SM electroweak gauge invariance.
One possibility is to extend the electroweak gauge group, for example,
by using the \TrTrOne group to exploit its peculiar
features~\cite{Singer:1980sw}. In this framework it has recently been
shown that one can obtain a type-II seesaw mechanism for Dirac
neutrinos~\cite{Valle:2016kyz, Addazi:2016xuh}.
One can also use a $B-L$ gauge extension with unconventional
charges for right handed neutrinos, which leads to Dirac neutrinos
obtained from type-I seesaw mechanism \cite{Ma:2014qra, Ma:2015raa,
  Ma:2015mjd}.
Alternatively one may stick to the simplest \SM gauge structure but
use extra symmetries implying a conserved lepton number, so as to
obtain Dirac neutrinos, as suggested in~\cite{Aranda:2013gga}.
One can also adopt extra-dimensional
  theory frameworks~\cite{extra}.

Here we focus on the possibility of requiring that neutrino
masses arise from a simple type-I seesaw mechanism within a
flavor--symmetric scenario. Moreover we will also require that the
existence of a viable dark matter particle arises from the same
symmetry which ensures that neutrinos do not acquire Majorana masses
and remain Dirac fermions.
In Sect.~\ref{sec:model-} we sketch in some detail the extended
particle content required to realize the non-Abelian flavor symmetry
of the model and demonstrate how the Dirac nature of neutrinos and the
smallness of their seesaw--induced masses follow from our non-Abelian
discrete flavor symmetry.
In Sect.~\ref{sec:numerical-scan} we present numerical predictions for
CP violation in terms of the scalar boson alignment patterns.
Towards the end of the paper, in Sect.~\ref{sec:dark-matter}, we
discuss the appearance of viable dark matter in this model and give a
brief discussion of its direct detection potential.
Finally we summarize our results in
Sect.~\ref{sec:summary-conclusions-}.


 \section{The Model }
\label{sec:model-}


Our model is based on the discrete flavor symmetry
$\Delta (27) \otimes Z_4$ where $Z_4$ is the cyclic group of order
four and $\Delta (27)$ is a discrete non-Abelian
  symmetry group isomorphic to $(Z_3 \otimes Z_3) \ltimes
  Z_3$. Before presenting the details of the model, we briefly
discuss the most relevant features of the $\Delta (27)$ group.
The $\Delta (27)$ group belongs to the general class of discrete
groups denoted by $\Delta (3 N^2)$, with $N$ being a positive
integer. The smallest member of $\Delta (3 N^2)$ is $\Delta (3)$ which
is nothing but the Abelian group $Z_3$. The next member is $\Delta
(12)$ which is isomorphic to the well--known group $A_4$.  The third
smallest member of the group is $\Delta (27)$ and has 27 elements
divided into 11 conjugacy classes \cite{ Ma:2006ip, Ma:2007wu,
  Ishimori:2010au, Aranda:2013gga}. 
It has nine singlet irreducible representations $\mathbf{1}_i$ ; $i =
1, \dots 9$ and two triplet irreducible representations $\mathbf{3}$
and $\mathbf{3'}$~\footnote{Here we denote the irreducible
  representations of $\Delta (27)$ as in~\cite{Aranda:2013gga},
  instead of the alternative ``two index'' notation used in
  \cite{Ishimori:2010au}. The two are related by : $\mathbf{1} \equiv
  \mathbf{1}_{(0,0)}$, $\mathbf{1'} \equiv \mathbf{1}_{(2,0)}$,
  $\mathbf{1''} \equiv \mathbf{1}_{(1,0)}$, $\mathbf{3} \equiv
  \mathbf{3}_{(0,1)}$, $\mathbf{3'} \equiv \mathbf{3}_{(0,2)}$. }. The
multiplication rules for $\Delta (27)$ are given by
\begin{eqnarray}
 \mathbf{3} \, \otimes \, \mathbf{3} & = & \mathbf{3'} \, \oplus \, \mathbf{3'} \, \oplus \, \mathbf{3'} \, ; \qquad \qquad 
 \mathbf{3} \, \otimes \, \mathbf{3'}  \, \, = \, \, \sum_{i = 1}^9  \, \mathbf{1}_i~.
 \label{muldel}
\end{eqnarray}
The particle content of our model along with the $\Delta (27) \otimes
Z_4$ charge assignments of the particles are as shown in Table
\ref{tab1}.

\begin{table}[h]
\begin{center}
\begin{tabular}{c c c || c c c}
  \hline \hline
  Fields           \hspace{1cm}        & $\Delta(27)$          \hspace{1cm}        &  $Z_4$                    \hspace{1cm}      & 
  Fields           \hspace{1cm}        & $\Delta(27)$          \hspace{1cm}        &  $Z_4$                                     \\
  \hline \hline
  $\bar{L}_e$      \hspace{1cm}        & $\mathbf{1}$          \hspace{1cm}        &  $\mathbf{z}^3$      \hspace{1cm}      &  
  $\nu_{e,R}$      \hspace{1cm}        & $\mathbf{1}$          \hspace{1cm}        &  $\mathbf{z}$                          \\
  $\bar{L}_\mu$    \hspace{1cm}        & $\mathbf{1''}$        \hspace{1cm}        &  $\mathbf{z}^3$      \hspace{1cm}      &
  $\nu_{\mu, R}$   \hspace{1cm}        & $\mathbf{1'}$         \hspace{1cm}        &  $\mathbf{z}$                          \\
  $\bar{L}_\tau$   \hspace{1cm}        & $\mathbf{1'}$         \hspace{1cm}        &  $\mathbf{z}^3$      \hspace{1cm}      &   
  $ \nu_{\tau,R}$  \hspace{1cm}        & $\mathbf{1''}$        \hspace{1cm}        &  $\mathbf{z}$                          \\
  $ l_{i,R} $      \hspace{1cm}        & $\mathbf{3}$          \hspace{1cm}        &  $\mathbf{z}$        \hspace{1cm}      & 
  $\bar{N}_{i,L}$  \hspace{1cm}        & $\mathbf{3}$          \hspace{1cm}        &  $\mathbf{z}^3$                        \\
  $N_{i,R}$        \hspace{1cm}        & $\mathbf{3'}$         \hspace{1cm}        &  $\mathbf{z}$         \hspace{1cm}     &                                      
                   \hspace{1cm}        &                       \hspace{1cm}        &                                             \\
  \hline
  $\Phi_i$         \hspace{1cm}        & $\mathbf{3'}$         \hspace{1cm}        &  $\mathbf{1}$             \hspace{1cm}      &
  $\chi_i$         \hspace{1cm}        & $\mathbf{3'}$         \hspace{1cm}        &  $\mathbf{1}$                              \\
  $\zeta$          \hspace{1cm}        & $\mathbf{1}$          \hspace{1cm}        &  $\mathbf{z}$        \hspace{1cm}      &               
  $\eta$           \hspace{1cm}        & $\mathbf{1}$          \hspace{1cm}        &  $\mathbf{z}^2$                        \\
    \hline
  \end{tabular}
\end{center}
\caption{ The $\Delta (27)$ and $Z_4$ charge assignments for leptons,
  the Higgs scalars ($\Phi_i$, $\chi_i$) and the dark matter sector
  scalars ($\zeta$ and $\eta$). Here $\mathbf{z}$ is the fourth root
  of unity, i.e. $\mathbf{z}^4 = 1$.}
  \label{tab1}
\end{table}
 
In Table \ref{tab1} $L_i = (\nu_i, l_i)^T$; $i = e, \mu, \tau$ are the
lepton doublets which also transform as singlets under $\Delta (27)$
and have charge $\mathbf{z}$ under $Z_4$. The $l_{i,R}$; $i = e, \mu,
\tau$ are the charged lepton singlets which transform as a
$\mathbf{3}$ under $\Delta (27)$ and have $Z_4$--charge $\mathbf{z}$.
Apart from the \sm fermions, the model also includes three
right--handed neutrinos $\nu_{i,R}$ transforming as singlets under the
\SM gauge group and as singlets under $\Delta (27)$, with charge
$\mathbf{z}$ under $Z_4$. We also add three gauge singlet Dirac fermions
$N_{i,L}, N_{i,R}$; $i = 1, 2, 3$ transforming as triplets of $\Delta
(27)$ and with charge $\mathbf{z}$ under $Z_4$, as shown in
Table.~\ref{tab1}.
 
\begin{figure}[!h]
\centering
\includegraphics[width=.6\textwidth]{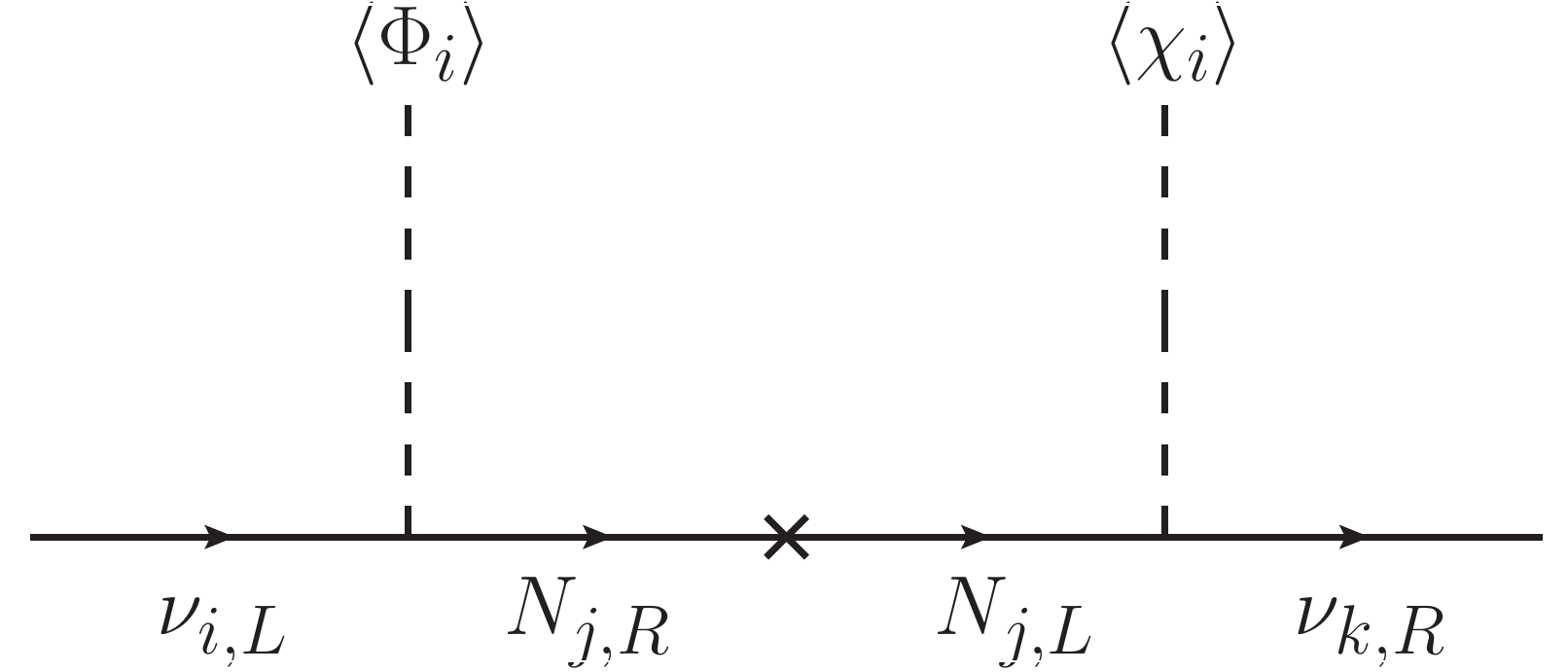}
\caption{The Dirac type-I seesaw mechanism.  $\Phi_i$ and $\chi_i$ are
  triplets under $\Delta(27)$. }
\label{feyn}
\end{figure}

In the scalar sector the $\Phi_i = (\phi^+_i, \phi^0_i)^T$; $i = 1, 2,
3$ transform as $\mathrm{SU(2)_L}$ doublets, as triplet under $\Delta
(27)$ and trivially under $Z_4$.
On the other hand the scalars $\chi_i$; $i = 1, 2, 3$ are gauge
singlets transforming as a triplet under $\Delta (27)$ and trivially
under $Z_4$. We also add two other gauge singlet scalars $\zeta$ and
$\eta$ both of which transform trivially under $\Delta (27)$ but carry
$Z_4$ charges $\mathbf{z}$ and $\mathbf{z}^2$ respectively. 
  Since $\mathbf{z}^2 = -1$, the field $\eta$ can be real and is taken to
  be real.
  We comment on the important role of the scalars $\zeta$ and $\eta$
  towards the end of the paper.
The \SM $ \otimes$ $\Delta (27) \otimes Z_4$ invariant Yukawa term for
the charged leptons is given by
 \begin{eqnarray}
  \mathcal{L}_{\rm{Yuk}, l}  & = & y_1 \, \left (\bar{L}_e \right)_1  \, \otimes \, \left [  \left( \begin{array}{c}  l_{e,R}   \\ l_{\mu, R} \\ l_{\tau, R} \end{array} \right)_3 \, \otimes  \left( \begin{array}{c} \Phi_1   \\ \Phi_2 \\ \Phi_3 \end{array} \right)_{3'}   \right]_1 
  \, + \, 
  y_2 \, \left (\bar{L}_\mu \right)_{1''} \, \otimes \,  \left [  \left( \begin{array}{c}  l_{e,R}   \\ l_{\mu, R} \\ l_{\tau, R} \end{array} \right)_3 \, \otimes  \left( \begin{array}{c} \Phi_1   \\ \Phi_2 \\ \Phi_3 \end{array} \right)_{3'}   \right]_{1'}
  \nonumber \\
  & + &  y_3 \, \left (\bar{L}_\tau \right)_{1'} \, \otimes \,  \left [  \left( \begin{array}{c}  l_{e,R}   \\ l_{\mu, R} \\ l_{\tau, R} \end{array} \right)_3 \, \otimes  \left( \begin{array}{c} \Phi_1   \\ \Phi_2 \\ \Phi_3 \end{array} \right)_{3'}   \right]_{1''}
  \label{lepyuk}
 \end{eqnarray}
 where $y_i$, $i = 1, 2, 3$, are the Yukawa couplings which, for
 simplicity, we take them to be real. After symmetry breaking the
 scalars acquire vacuum expectation values (vevs) $\vev{ \Phi_i} =
 v_i$; $i = 1, 2, 3$ so the charged lepton mass matrix is given by
  \begin{eqnarray}
   M_{l} = \left( 
\begin{array}{ccc}
y_1 v_1    & y_1 v_2           &   y_1 v_3 \\
y_2 v_1    & \omega y_2 v_2    &   \omega^2 y_2 v_3 \\
y_3 v_1    & \omega^2 y_3 v_2  &   \omega y_3 v_3  \\  
\end{array}
\right)~.
   \label{glepmass}
  \end{eqnarray}
  The corresponding Yukawa term, relevant for generating masses for
  the neutrinos and the heavy neutral fermions $N_{L}, N_{R}$ is given
  by
 \begin{eqnarray}
  \mathcal{L}_{\rm{Yuk}, \nu}  & = & a_1 \, \left (\bar{L}_e \right)_1  \, \otimes \, \left [  \left( \begin{array}{c} \tilde{\Phi}_1   \\ \tilde{\Phi}_2 \\ \tilde{\Phi}_3 \end{array} \right)_{3} \, \otimes \,  \left( \begin{array}{c}  N_{1,R}   \\ N_{2, R} \\ N_{3, R} \end{array} \right)_{3'}  \right]_1 
  \, + \, 
  a_2 \, \left (\bar{L}_\mu \right)_{1''}  \, \otimes \, \left [  \left( \begin{array}{c} \tilde{\Phi}_1   \\ \tilde{\Phi}_2 \\ \tilde{\Phi}_3 \end{array} \right)_{3} \, \otimes \,  \left( \begin{array}{c}  N_{1,R}   \\ N_{2, R} \\ N_{3, R} \end{array} \right)_{3'}  \right]_{1'} 
  \nonumber \\
  & + &   a_3 \, \left (\bar{L}_\tau \right)_{1'}  \, \otimes \, \left [  \left( \begin{array}{c} \tilde{\Phi}_1   \\ \tilde{\Phi}_2 \\ \tilde{\Phi}_3 \end{array} \right)_{3} \, \otimes \,  \left( \begin{array}{c}  N_{1,R}   \\ N_{2, R} \\ N_{3, R} \end{array} \right)_{3'}  \right]_{1''} 
  \, + \,
  b_1 \,   \left [ \left( \begin{array}{c}  \bar{N}_{1,L}   \\ \bar{N}_{2, L} \\ \bar{N}_{3, L} \end{array} \right)_{3} \, \otimes \, \left( \begin{array}{c} \chi_1   \\ \chi_2 \\ \chi_3 \end{array} \right)_{3'} \right]_1  \, \otimes \, \left(\nu_{e,R} \right)_1
  \nonumber \\
  & + &
   b_2 \,   \left [ \left( \begin{array}{c}  \bar{N}_{1,L}   \\ \bar{N}_{2, L} \\ \bar{N}_{3, L} \end{array} \right)_{3} \, \otimes \, \left( \begin{array}{c} \chi_1   \\ \chi_2 \\ \chi_3 \end{array} \right)_{3'} \right]_{1''}  \, \otimes \, \left(\nu_{\mu,R} \right)_{1'}
   \, + \, 
    b_3 \,   \left [ \left( \begin{array}{c}  \bar{N}_{1,L}   \\ \bar{N}_{2, L} \\ \bar{N}_{3, L} \end{array} \right)_{3} \, \otimes \, \left( \begin{array}{c} \chi_1   \\ \chi_2 \\ \chi_3 \end{array} \right)_{3'} \right]_{1'}  \, \otimes \, \left(\nu_{\tau,R} \right)_{1''}
    \nonumber \\
    & + &     M \,   \left( \begin{array}{c}  \bar{N}_{1,L}   \\ \bar{N}_{2, L} \\ \bar{N}_{3, L} \end{array} \right)_{3} \,
    \otimes \,\left( \begin{array}{c}  N_{1,R}   \\ N_{2, R} \\ N_{3, R} \end{array} \right)_{3'} 
  \label{neutyuk}
 \end{eqnarray}
 where $a_i, b_i$; $i = 1, 2, 3$ are the Yukawa couplings which are
 taken to be real. The parameter $M$ is the gauge and
 flavor--invariant mass term for the heavy leptons. After symmetry
 breaking the scalars $\chi_i$ also acquire vevs $\vev{ \chi_i} =
 u_i$; $i = 1, 2, 3$.
 The $6\times 6$ mass matrix for the neutrinos and the heavy fermions
 in basis $( \bar{\nu}_{e,L}, \bar{\nu}_{\mu,L}, \bar{\nu}_{\tau,L},
 \bar{N}_{1,L}, \bar{N}_{2,L}, \bar{N}_{3,L} )$ and $( \nu_{e,R},
 \nu_{\mu,R}, \nu_{\tau,R}, N_{1,R}, N_{2,R}, N_{3,R} )^T$ is given by
  \begin{eqnarray}
   M_{\nu, N} = \left( 
\begin{array}{cccccc}
0          & 0                 &  0                  & a_1 v_1    & a_1 v_2           &   a_1 v_3             \\
0          & 0                 &  0                  & a_2 v_1    & \omega a_2 v_2    &   \omega^2 a_2 v_3    \\
0          & 0                 &  0                  & a_3 v_1    & \omega^2 a_3 v_2  &   \omega a_3 v_3       \\ 
b_1 u_1    & b_2 u_1           &  b_3 u_1            & M          & 0                 &  0                     \\
b_1 u_2    & \omega b_2 u_2    & \omega^2 b_3 u_2    & 0          & M                 &  0                     \\
b_1 u_3    & \omega^2 b_2 u_3  & \omega b_3 u_3      & 0          & 0                 &  M                     \\
\end{array}
\right)
   \label{gneutmass}
  \end{eqnarray}
  The invariant mass term $M$ for the heavy leptons $N_L, N_R$ can be
  naturally much larger than the symmetry breaking scales appearing in
  the off-diagonal blocks, i.e.  $ M \gg v_i, u_i$. In this limit the
  mass matrix in (\ref{gneutmass}) can be easily block diagonalized by
  the perturbative seesaw diagonalization method given
  in~\cite{Schechter:1981cv}. The resulting $3 \times 3$ mass matrix
  for light neutrinos can be viewed as the Dirac version of the well
  known type-I seesaw mechanism. The above mass generation mechanism
  can also be represented diagramatically as shown in
  Fig. \ref{feyn}.

The $3 \times 3$ light neutrino mass matrix along with the charged
lepton mass matrix (\ref{glepmass}) have enough free parameters to
account for all the observed mass and mixing parameters in the lepton
sector.
As has been discussed in several previous works~\cite{ Ma:2006ip,
  Ma:2007wu, Varzielas:2015aua, Aranda:2013gga, Hernandez:2016eod,
  Vien:2016tmh}, for $\Delta (27)$ we focus on the vev alignment $v_1
= v_2 = v_3 = v$ and $u_1 = u_2 = u_3 = u$ as a reference case. Taking
this ``double alignment'' limit for the vevs of the scalars the
charged lepton mass matrix $M_l$ is given by 
   \begin{eqnarray}
   M_{l} = v \, \left( 
\begin{array}{ccc}
y_1     & y_1            &   y_1           \\
y_2     & \omega y_2     &   \omega^2 y_2  \\
y_3     & \omega^2 y_3   &   \omega y_3    \\  
\end{array}
\right)~,
   \label{alglepmass}
  \end{eqnarray}
  and can be easily diagonalized from right by the familiar ``magic
  matrix'' $U_\omega$ given by
  \begin{eqnarray}
  U_{\omega} = \dfrac{1}{\sqrt{3}} \, \left( 
\begin{array}{ccc}
1     & 1           &  1           \\
1     & \omega      &  \omega^2   \\
1     & \omega^2    &  \omega     \\  
\end{array}
\right)~.
   \label{magmat}
  \end{eqnarray}
 This leads to
 \begin{eqnarray}
  M_{l} \, . \, U^\dagger_{\omega} & = &  \left( 
\begin{array}{ccc}
 \sqrt{3} v y_1   & 0                  &  0                \\
0                 & \sqrt{3} v y_2     &  0                 \\
0                 & 0                  &  \sqrt{3} v y_3     \\  
\end{array}
\right)~.
\label{alglepdig}
 \end{eqnarray}

 Likewise, the neutral fermion mass matrix $M_{\nu, N}$ in the above
 alignment limit is given by
   \begin{eqnarray}
   M_{\nu, N} = \left( 
\begin{array}{cccccc}
0          & 0                 &  0                  & a_1 v    & a_1 v           &   a_1 v             \\
0          & 0                 &  0                  & a_2 v    & \omega a_2 v    &   \omega^2 a_2 v    \\
0          & 0                 &  0                  & a_3 v    & \omega^2 a_3 v  &   \omega a_3 v       \\ 
b_1 u      & b_2 u             &  b_3 u              & M          & 0                 &  0                     \\
b_1 u      & \omega b_2 u      & \omega^2 b_3 u      & 0          & M                 &  0                     \\
b_1 u      & \omega^2 b_2 u    & \omega b_3 u        & 0          & 0                 &  M                     \\
\end{array}
\right)
   \label{algneutmass}
  \end{eqnarray}
  As mentioned before the invariant mass term $M$ for the heavy
  fermions $N_L, N_R$ is naturally expected to be much larger than the
  symmetry breaking scale i.e. $v, u \ll M$. In such limit the mass
  matrix in (\ref{algneutmass}) can be easily block--diagonalized. The
  resulting $3 \times 3$ mass matrix for the light neutrinos assuming
  such simplest alignment is given by
\begin{eqnarray}
  M_{\nu}  & = &  \dfrac{u \, v }{M} \, \left( 
\begin{array}{ccc}
 3 a_1 b_1       & 0                 &  0                \\
0                & 0                 &  3 a_2 b_3        \\
0                & 3 a_3 b_2         &  0                 \\  
\end{array}
\right)
\label{algneutmass}
 \end{eqnarray}
 Clearly the light neutrino mass matrix in Eq.~(\ref{algneutmass}) is
 inconsistent with the current neutrino oscillation data
 \cite{Forero:2014bxa} and needs to be modified.
 
 In order to obtain a realistic light neutrino mass spectrum one must
 generalize the above vev--alignment pattern i.e.  $v_1 = v_2 = v_3 =
 v$ and $u_i = u_j = u, u_k \neq u$ where $i,j,k = 1, 2, 3$.
 Thus in our generalized ansatz we keep the alignment for the
 isodoublet scalar vevs unchanged, but modify the isosinglet scalars
 vev alignment.
 Such a generalization is not unfounded since the scalar sector of our
 model is much richer than that characterizing the simpler case of
 only one type of scalars transforming as $\Delta (27)$ triplets, 
 discussed in \cite{ Ma:2006ip, Ma:2007wu, Varzielas:2015aua,
   Aranda:2013gga}.
 In contrast to previous models we have two different types of scalars
 namely $\Phi_i$ and $\chi_i$ both transforming as triplets under
 $\Delta (27)$. The resulting scalar potential is rich enough to allow
 for other possible vev alignments to be realized.

 We find that any of the three possible choices namely $u_1 = u_2 = u,
 u_3 \neq u$; $u_2 = u_3 = u, u_1 \neq u$; $u_1 = u_3 = u, u_2 \neq u$
 can give realistic neutrino mass matrices. 
 However, for definiteness and to avoid unnecessary repetition
 henceforth we focus on the
 choice $u_1 = u_3 = u, u_2 \neq u$.  Towards the end of the discussion we
 will comment on the similarities and differences in results for other
 possibilities.

 Since we have kept the vev alignment for the $\Phi_i$ fields
 unchanged it follows that the charged lepton mass matrix
 Eq.~(\ref{alglepmass}) also remains unchanged. As a result it can
 still be diagonalized by a ``magic'' rotation from the right as shown
 in (\ref{alglepdig}). The $6 \times 6$ neutral fermion mass matrix
 now becomes
  \begin{eqnarray}
   M_{\nu, N} = \left( 
\begin{array}{cccccc}
0          & 0                 &  0                  & a_1 v      & a_1 v             &   a_1 v             \\
0          & 0                 &  0                  & a_2 v      & \omega a_2 v      &   \omega^2 a_2 v    \\
0          & 0                 &  0                  & a_3 v      & \omega^2 a_3 v    &   \omega a_3 v       \\ 
b_1 u      & b_2 u             &  b_3 u              & M          & 0                 &  0                     \\
b_1 u_2    & \omega b_2 u_2    & \omega^2 b_3 u_2    & 0          & M                 &  0                     \\
b_1 u      & \omega^2 b_2 u    & \omega b_3 u        & 0          & 0                 &  M                     \\
\end{array}
\right)
\label{neutmass}
\end{eqnarray}
As before this mass matrix can be block--diagonalized in the
approximation $v, u, u_3 \ll M$. The resulting light three--neutrino
mass matrix is
\begin{eqnarray}
  M_{\nu}  & = &  \dfrac{ v }{M} \, \left( 
\begin{array}{ccc}
a_1 b_1 (2 u + u_2)                     & a_1 b_2 (u + \omega^2 u + \omega u_2)   & a_1 b_3 (u + \omega u + \omega^2 u_2) \\
a_2 b_1 (u + \omega^2 u + \omega u_2)   & a_2 b_2 (u + \omega u + \omega^2 u_2)   & a_2 b_3 (2 u + u_2)                   \\
a_3 b_1 (u + \omega u + \omega^2 u_2)   & a_3 b_2 (2 u + u_2)                     & a_3 b_3 (u + \omega^2 u + \omega u_2) \\  
\end{array}
\right)
\label{neutmass}
 \end{eqnarray}


 \section{CP violation}
\label{sec:numerical-scan} 

 
The neutrino mass matrix in Eq.~(\ref{neutmass}) can be diagonalized
numerically and leads to neutrino masses and mixing angles consistent
with neutrino oscillation experiments \cite{Forero:2014bxa} as well as
cosmological limits \cite{Ade:2015xua}. 
Here we present our numerical results for CP violation in this model.
Notice that from the beginning, we have assumed real Yukawa
couplings. If we also take a real scalar potential, leptonic CP
violation must arise solely by the complex nature of the $\Delta (27)$
flavor symmetry.
Indeed, one finds that, with our generalized alignment the resulting
neutrino mass matrix (\ref{neutmass}) leads to no CP violation and in
terms of standard parametrization of neutrino mixing matrix
\cite{Forero:2014bxa}, one has $\delta_{CP} = 0, \pm \pi$ for the CP
phase.
The latter implies that the Jarlskog invariant $J_{CP}$, which in the
standard PDG parametrization~\footnote{For a recent discussion of
  fermion mixing parametrizations see~\cite{Rodejohann:2011vc}.} is
given by $$ J_{CP} = \frac{1}{8}\, \sin{2 \theta_{12}} \sin{2
  \theta_{23}} \sin{2 \theta_{13}} \cos{\theta_{13}
  \sin{\delta_{CP}}},$$ vanishes.  
 
\begin{figure}[!h]
\centering
\includegraphics[width=.45\textwidth]{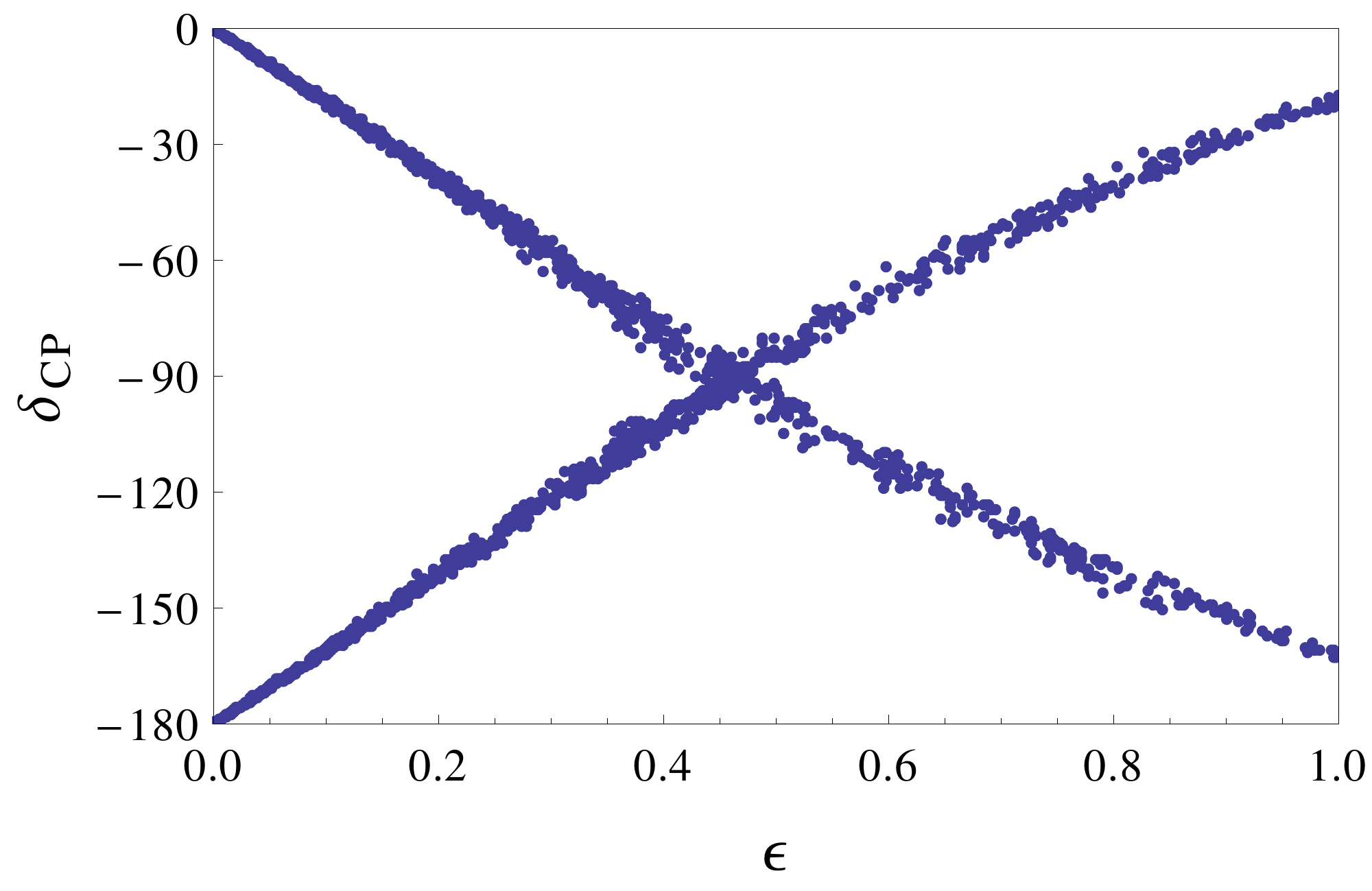}
\includegraphics[width=.45\textwidth]{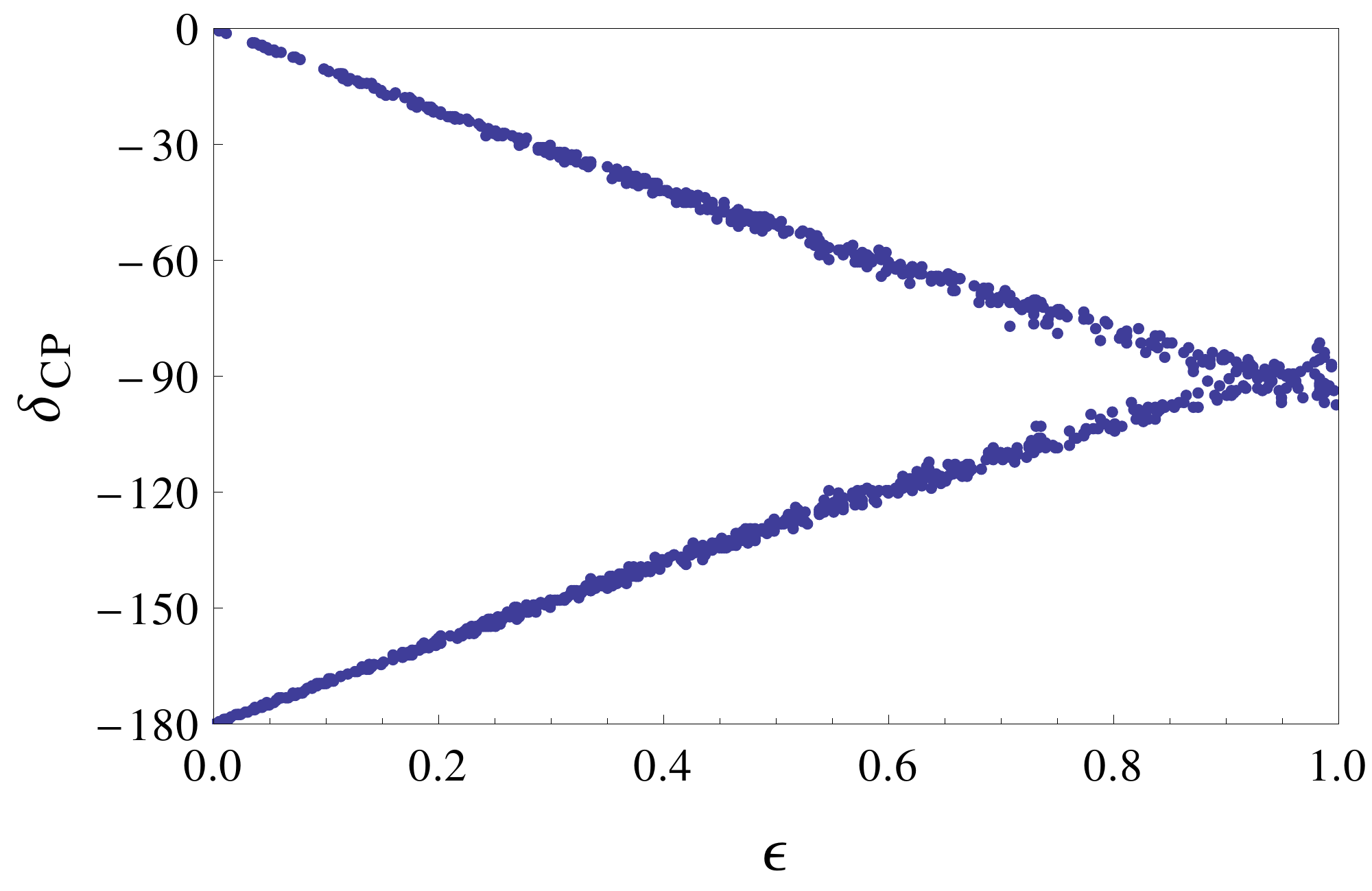}
\caption{ Leptonic CP violation phase $\delta_{CP}$
    versus $\epsilon$, the deviation from the reference alignment. For
    the left panel we have taken $\alpha = 1.2$ whereas in the right
    panel the $\alpha = 2.5$ is taken. See text. }
\label{fig1}
\end{figure}

\begin{figure}[htb]
\includegraphics[width=.45\textwidth]{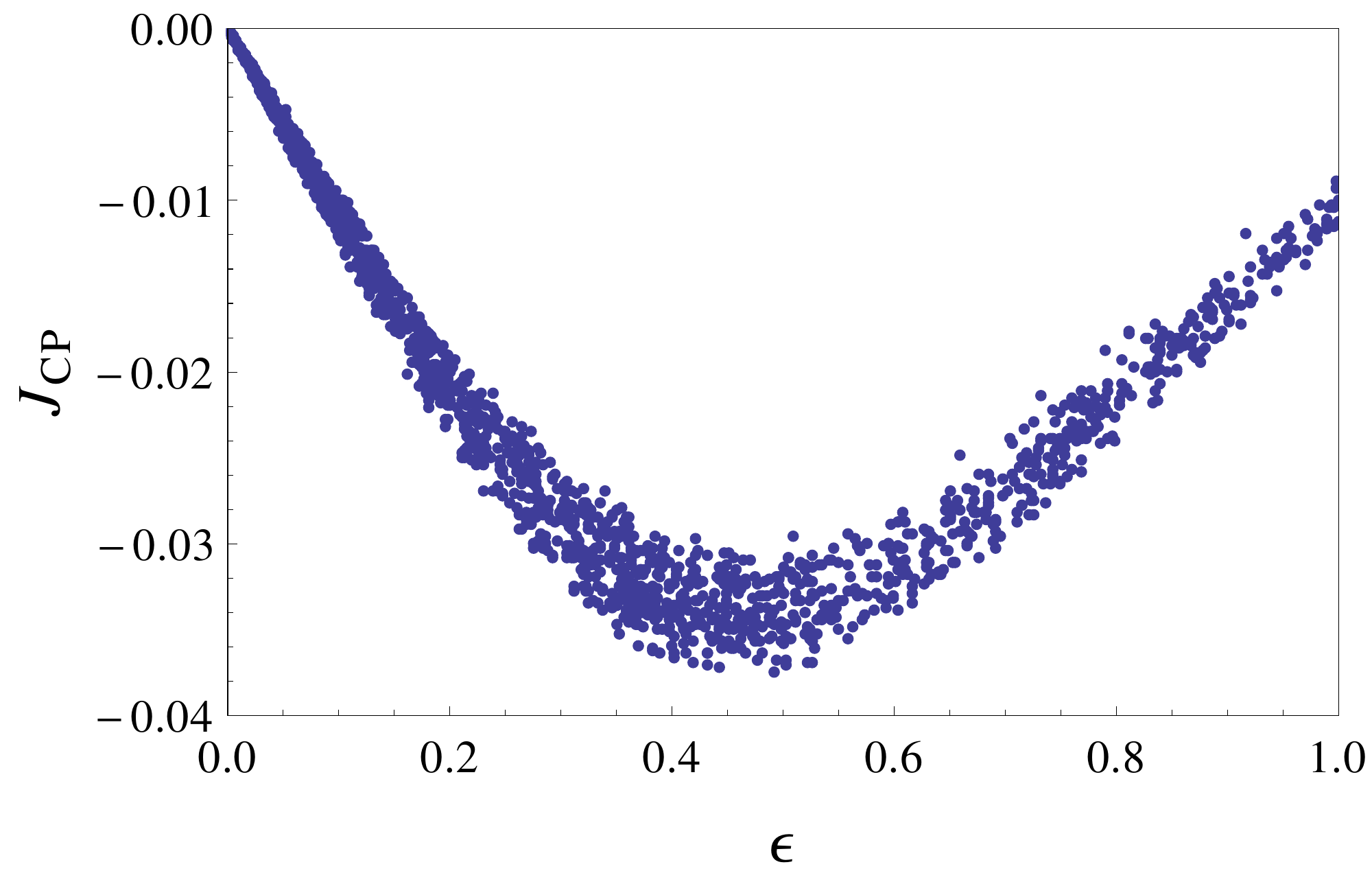}
\includegraphics[width=.45\textwidth]{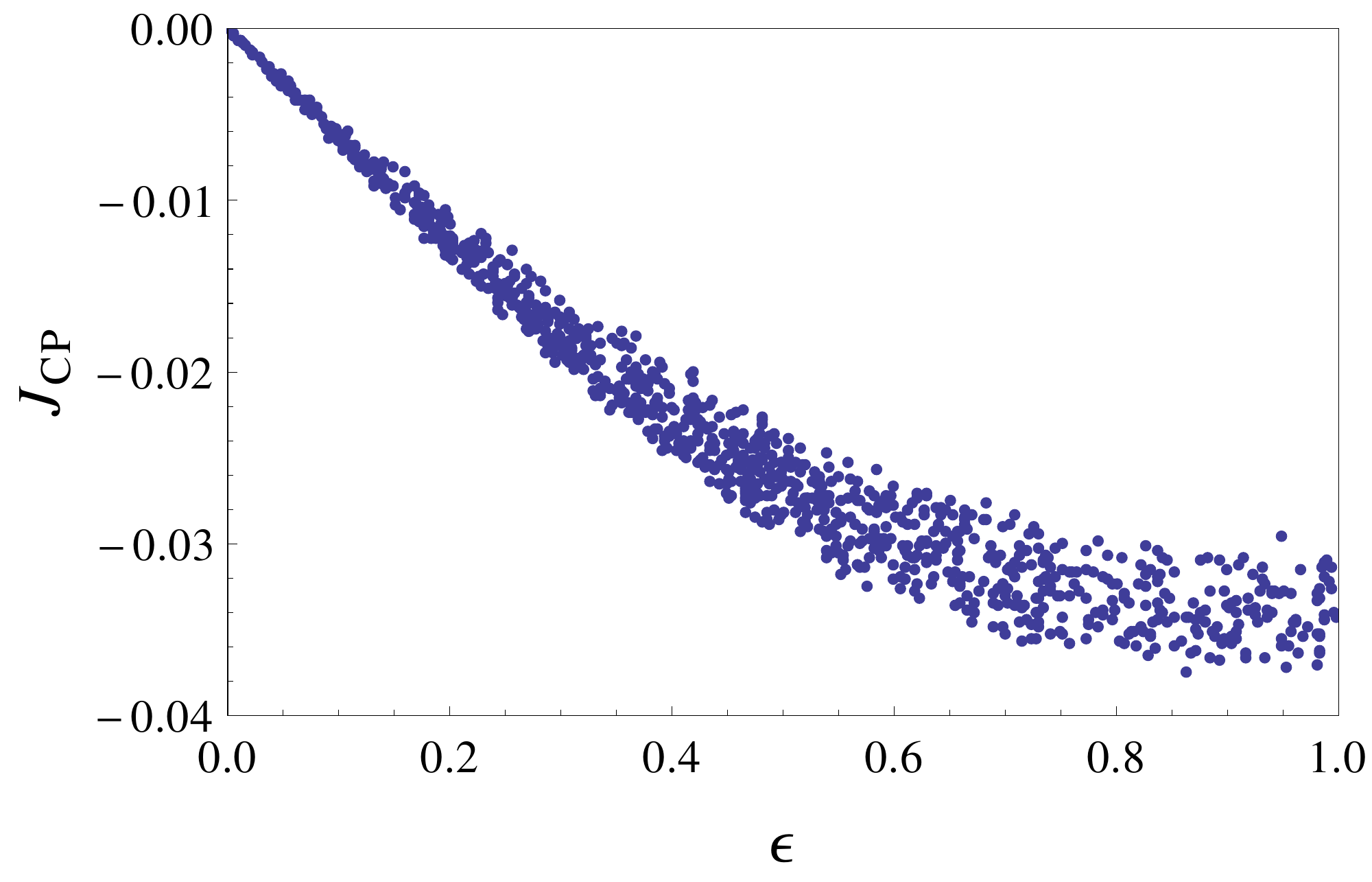}
\caption{The Jarlskog invariant $J_{CP}$ versus the deviation from
  alignment $\epsilon$.  The range of variation of other free
  parameters is the same as in \ref{fig1}. For the left panel we have
  taken $\alpha = 1.2$ whereas in the right panel we take $\alpha =
  2.5$. }
\label{fig2}
\end{figure}

Recent experimental results have predicted a slight preference for
$\delta_{CP} \neq 0, \pm \pi$ implying CP violation in lepton sector
\cite{Forero:2014bxa}. If this indeed is the case then one must
consider deviations from the generalized alignment limit. For example,
if we consider small deviation of the type
$u_1 = u, u_3 = u (1 + \epsilon), u_2 = u (1 + \alpha)$ then finite CP
violation can indeed be generated even for real $\epsilon$ and
$\alpha$, as shown in Fig \ref{fig1} and \ref{fig2}.  The source of CP
violation can be traced to the complex parameter $\omega$, where
$\omega$ is cube root of unity with $\omega^3 = 1$.

As can be seen from the figures, when $\epsilon = 0$, $\delta_{CP} =
0, - \pi$ and $J_{CP} = 0$ implying no CP violation. As we deviate
from our reference alignment limit CP violation is generated with
$J_{CP} \neq 0$. The magnitude of the CP violation parameter is
directly proportional to the deviation $\epsilon$ from the alignment
limit as well as the parameter $\alpha$ which measures the deviation
of $u_2$ from $u_1$ i.e. $u_2 = u_1 (1 + \alpha)$.
In plotting Fig.~\ref{fig1} we have randomly varied all other free
    parameters, namely the vevs and Yukawa couplings. All the Yukawa
    couplings are varied between $-0.5$ to $0.5$, the $u_1$ vev is
    varied between $700$ to $800$ GeV and the $u_2$ vev is taken to be
    $u_1 (1 + \alpha)$.  

    For a given value of $\alpha$ the magnitude of CP violation is
    directly correlated to $\epsilon$ as is clear from Fig. \ref{fig1}
    and Fig. \ref{fig2}.  In Fig. \ref{fig1} we show the deviation of
    $\delta_{CP}$ with respect to $\epsilon$ for fixed values of
    $\alpha$.
The dependence of the Jarlskog invariant $J_{CP}$ with respect to
$\epsilon$, for fixed values of $\alpha$, is shown in Fig. \ref{fig2}.
For the left panel of both figures, we have fixed $\alpha = 1.2$ while
for right panel we took $\alpha = 2.5$. As is clear from a comparison
of the two panels, the magnitude of CP violation not only depends on
$\epsilon$ but also on the value of $\alpha$. 
For smaller values of $\alpha$ the deviation is sharper than for
larger values.  In the left panels of the two figures, where a
relatively smaller value of $\alpha$ is taken, the $\delta_{CP}$ as
well as $J_{CP}$ changes rapidly with $\epsilon$ and maximal CP
violation corresponding to $\delta_{CP} = -\pi/2$ is obtained for
$\epsilon \approx 0.45$.
Further increase in $\epsilon$ values results in decrease in CP
violation as can be inferred from the decreasing value of $J_{CP}$ in
Fig \ref{fig2}. The $J_{CP}$ eventually falls back to zero with
$\delta_{CP} = 0, -\pi/2$, when $\epsilon = \alpha$ which again
corresponds to the reference alignment with $u_3$ now being equal to
$u_2$.
In the right panels of Fig. \ref{fig1} and Fig. \ref{fig2}, the
$\delta_{CP}$ and $J_{CP}$ are plotted with respect to $\epsilon$ for
a fixed values of $\alpha = 2.5$. The nature of the departures of both
$\delta_{CP}$ and $J_{CP}$ is similar to what is seen in the left
panels, but now the slope of the deviation is smaller. For $\alpha =
2.5$ maximal CP violation is achieved for higher value of $\epsilon
\approx 0.95$. Just like for the left panels, further increase in
$\epsilon$ beyond $0.95$ leads to decrease in CP violation with the
case of no CP violation i.e. $J_{CP} = 0$ with $\delta_{CP} = 0,
-\pi/2$ again achieved for $\epsilon = \alpha$ corresponding to the
alignment $u_3 = u_2$.
Notice that, although here we are presenting results only for positive
values of $\epsilon$ we mention that negative values of $\epsilon$ are
equally viable.  If we take $\epsilon < 0$ then the essential features
of Fig. \ref{fig1} and \ref{fig2} are reproduced but for positive
values of $\delta_{CP}$ and $J_{CP}$. This means that as $\epsilon$
deviates more and more from zero on the negative side, both
$\delta_{CP}$ and $J_{CP}$ start deviating more and more from the CP
conserving case but along the positive direction. Again the departure
depends also on the value of $\alpha$ with smaller values of $\alpha$
leading to sharper deviation with respect to $\epsilon$.

Finally before closing this section let us briefly remark on other
possible alignment choices, e.g.  $u_1 = u, u_2 = u (1 + \epsilon'),
u_3 = u (1 + \alpha') $ where $\epsilon'$ and $\alpha'$ parametrize
the deviations of $u_2, u_3$ from $u$, respectively. 
As in the previous case, here also for the case of perfect alignment
i.e. for $\epsilon' = 0$ we have no CP violation with $\delta_{CP} =
0, \pm \pi$ and $J_{CP} = 0$. Also as before when $\epsilon'$ deviates
from zero in either direction we generate CP violation. However,
unlike the previous case, the nature of the correlation in this case
is different, since for $\epsilon' > 0$ both $\delta_{CP}$ and
$J_{CP}$ acquire positive values, whereas for $\epsilon' < 0$ both
$\delta_{CP}, J_{CP} < 0 $. This behaviour is opposite to that
found in previous, case where for $\epsilon > 0$ we had $\delta_{CP},
J_{CP} < 0 $ and for $\epsilon < 0$ we had $\delta_{CP}, J_{CP} > 0
$. Apart from this, other features of the previous case like the
dependence on $\epsilon'$ and $\alpha'$ are qualitatively realized in
this case also.  
Finally, for the third alignment choice i.e. $u_2 = u, u_3 = u (1 +
\epsilon'), u_1 = u (1 + \alpha')$ the qualitative nature of CP
violation with respect to alignment deviation is essentially the same
as shown in Fig.~\ref{fig1} and Fig.~\ref{fig2}. To avoid unnecessary
repetition we refrain from discussing these two alignment choices in
more detail.


\section{WIMP scalar dark matter candidate}
\label{sec:dark-matter}

\begin{figure}[!h]
\centering
\includegraphics[width=.5\textwidth]{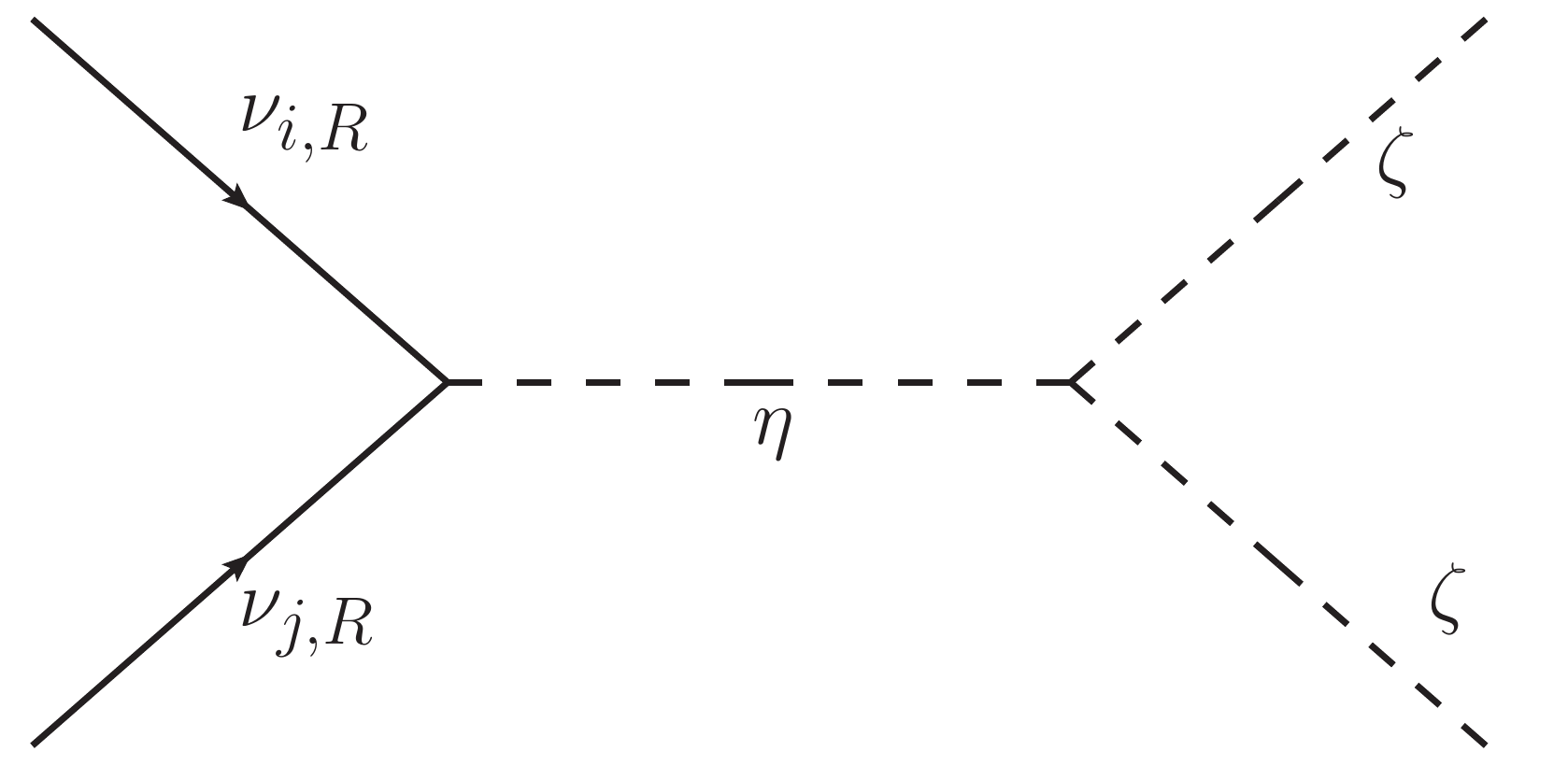}
\caption{Interaction between the dark matter candidate $\zeta$ and
  the right handed neutrinos, mediated by the exchange of the scalar $\eta$. }
\label{fig4}
\end{figure}

Here we recall the dark matter features of the model, which employs
similar ingredients as the simplest prototype model considered
in~\cite{Chulia:2016ngi}.
In this section we briefly consider the role of the scalars $\zeta$
and $\eta$, which are singlets under the \SM gauge group, transform
trivially under $\Delta(27)$, but carry $Z_4$ lepton quarticity
charges $\mathbf{z}$ and $\mathbf{z}^2$ respectively.
If $\zeta$ and $\eta$ are removed, the Lagrangian of the model
presents a larger symmetry associated to \SM
$\, \otimes \, \Delta (27) \, \otimes \, U(1)$ where $U(1)$ is a
continuous global symmetry which may be interpreted as a generalized
global lepton number.
However, in the presence of the scalars $\zeta$ and $\eta$ one can
write following $Z_4$ invariant terms in the scalar potential
\begin{eqnarray}
 \eta^2,\, \, \eta \, \zeta^2, \, \, \eta^4, \, \, \zeta^4, \, \,  \eta^2 \, \zeta^* \, \zeta \quad + \quad \rm{h.c.} 
 \label{z4invpot}
\end{eqnarray}
Notice that all these terms are $Z_4$ invariant but break the global
$U(1)$ invariance so the remaining family symmetry group is just
$\Delta (27) \otimes Z_4$~\footnote{We do not bother writing the other
  scalar potential terms which are also invariant under the global
  U(1).}.
On the other hand note that the field $\eta$ also couples to the right
handed neutrinos through a $Z_4$ invariant term
\begin{eqnarray}
  \bar{\nu}^c_{i,R} \, \nu_{j,R} \, \eta \, \, + \, \, \rm{h.c.}
 \label{z4invyuk}
\end{eqnarray}
Since this Yukawa coupling is only $Z_4$ invariant, it breaks the
continuous $U(1)$ symmetry. Due to the couplings of $\eta$ to the
scalar $\zeta$ in Eq.~\ref{z4invpot} and to right handed neutrinos as
in Eq.~\ref{z4invyuk}, the latter also couple to $\zeta$ as shown in
Fig. \ref{fig4}.

Note that the flavor symmetry $\Delta (27)$ breaks spontaneously when
the $\Delta (27)$ triplet scalars $\Phi_i$ and $\chi_i$ acquire
nonzero vevs. However, since neither $\Phi_i$ nor $\chi_i$ carries the
$Z_4$ charge, and $\zeta$ and $\eta$ are assumed not to acquire any
vev, one finds that the $Z_4$ remains unbroken.
This implies that the neutrinos retain their Dirac nature, since
Majorana mass terms are forbidden by the unbroken $Z_4$.

As a result one finds that the field $\zeta$ can act as a stable
particle and hence a potential candidate for the cosmological dark
matter. This implies that there is no term of the form
$\zeta \rho_i \rho_j$ or of the form $\zeta \psi_i \psi_j$, where
$\rho_i, \rho_j$ stand for other scalars and $\psi_j$, $\psi_i$ denote
generic fermions. Thus, the residual $Z_4$ symmetry responsible for
the Dirac nature of neutrinos also ensures the stability of the
$\zeta$ making it a potentially viable dark matter candidate.

 \begin{figure}[!h]
\centering
\includegraphics[width=.8\textwidth]{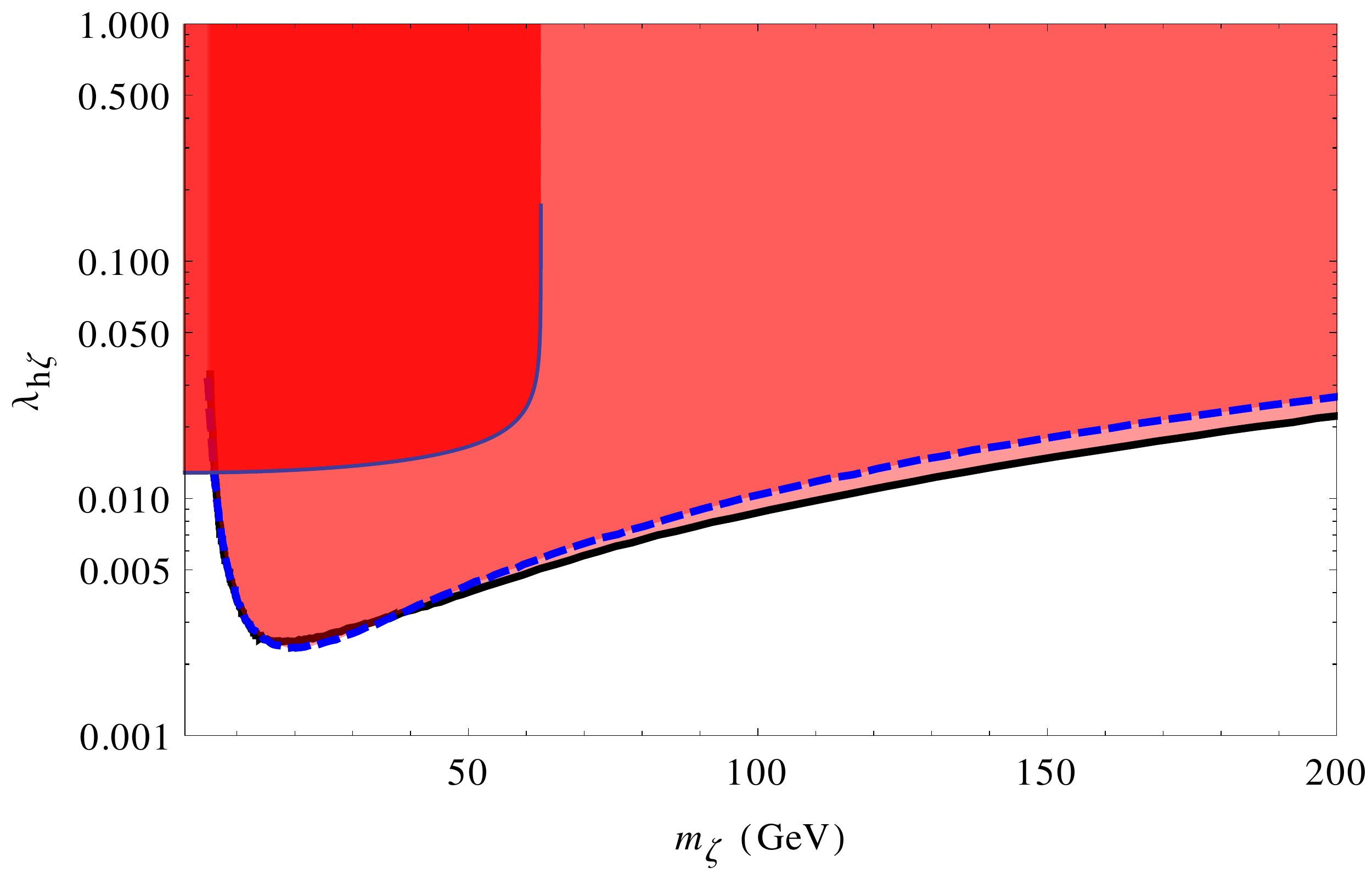}
\caption{The experimental sensitivity of our WIMP scalar dark matter
  candidate to invisible Higgs decay and direct detection. The light
  shaded region is ruled out by LUX (black continuous line) \cite{Akerib:2016vxi} and PandaX (blue dashed line) \cite {Tan:2016zwf} data whereas the dark shaded region is ruled out by the bound on the Higgs invisible decay width from the LHC~\cite{Aad:2015txa}. }
\label{cons}
\end{figure}

Although $\zeta$ is stable, and without direct tree level coupling to
fermions, due to the model symmetry, it still interacts with other
scalars through quartic potential terms of the type
$\zeta^* \zeta \rho^\dagger_i \rho_j$ and also couples to right handed
neutrinos through exchange of $\eta$ as shown in Fig. \ref{fig4}.
These terms imply that two dark matter particles can annihilate into
two other scalars, potentially leading to the correct relic density
for dark matter~\cite{Boucenna:2011tj, Ma:2015mjd}.  Also, the dark
matter interaction with the Higgs ($h$)\footnote{We denote the $125$
  GeV scalar discovered at LHC in 2012 as the Higgs. In our model it
  will be an admixture of the scalars $\Phi_i$ and $\chi_i$.} can be
used to detect it by experiments searching for nuclear
recoil~\cite{Ma:2015mjd} induced by Higgs boson exchange.
Moreover, if the dark matter mass obeys $m_\zeta < m_h/2$ then it can
lead to invisible decay of Higgs. Both nuclear recoil experiments such
as LUX \cite{Akerib:2016vxi} and PandaX \cite{Tan:2016zwf} as well as LHC searches for invisibly
decaying Higgs boson~\cite{Aad:2015txa, Khachatryan:2014jba} lead to
stringent constraints on the Higgs dark matter coupling as shown in
Fig.~\ref{cons}. In plotting Fig.~\ref{cons}, we have
  taken the constraints from the latest ATLAS searches for invisible
  Higgs decays~\cite{Aad:2015txa}, since the ATLAS constraint is more
  stringent than that of CMS~\cite{Khachatryan:2014jba}. Concerning
  constraints from nuclear recoil experiments, the LUX~\cite{Akerib:2016vxi}  and PandaX~\cite{Tan:2016zwf} experimental
  constraints are taken, assuming that the nucleon
  Higgs coupling and the nucleon mass parameters are the same as in
  ~\cite{Chulia:2016ngi}. Our treatment for dark matter constraints
  follows closely Ref.~\cite{Chulia:2016ngi} which should be consulted
  for further details.  Thus $\zeta$ realizes a ``Higgs portal'' dark
matter scenario. This type of dark matter, charged under a given
discrete symmetry, has been previously studied in several papers and
shown to provide a viable dark matter scenario \cite{Boucenna:2011tj,
  Ma:2015mjd, Cline:2013gha, Feng:2014vea}.
Another implication of our model is the conservation of the $Z_4$
charge in the presence of \lnv~\cite{Chen:2012jg, Heeck:2013rpa}. 
The fact that $\eta$ is a real scalar field which couples to right
handed neutrinos, means that its decay to two neutrinos or two
antineutrinos would potentially generate a lepton asymmetry in the
Universe. The possibility of leptogenesis with a conserved $Z_4$
lepton number has indeed been pointed out
in~\cite{Heeck:2013vha}. Clearly this scenario deserves more work.

\section{Discussion and Summary }
\label{sec:summary-conclusions-}

We have suggested a simple flavor model based on the $\Delta (27)$
group, in which the light neutrinos are Dirac fermions and the
smallness of their masses results from a type-I seesaw mechanism.
  Leptonic CP violation is related to the pattern of flavor symmetry
  breaking, described through the Higgs vacuum expectation values
  alignment, as shown in Figs.~\ref{fig1} and \ref{fig2} above.
  The scheme naturally leads to a WIMP dark matter candidate which is
  made stable by the same discrete lepton number $Z_4$ symmetry which
  makes neutrinos to be Dirac particles.
  In short, dark matter stability emerges from the lepton quarticity
  which also ensures the Dirac nature of neutrinos.
  A detailed study of its discovery potential in direct and indirect
  detection experiments will be presented elsewhere. Before closing
  let us also mention that our model can easily be generalized by
  including vector--like quarks, so as to accommodate the recent
  diphoton hint seen by the ATLAS and CMS collaborations. It would be
  identified with one of the scalars in the $\chi$ multiplet, very
  much along the lines of Refs.~\cite{Bonilla:2016sgx,Modak:2016ung}.
  In this paper we have discussed leptons only.  Quarks can be
  introduced in a trivial way as flavor singlets, along with a new
  Higgs scalar multiplet. 
   This Higgs scalar can be forbidden to couple with leptons by an additional $Z_2$
   symmetry in a way akin to the lepton specific two Higgs doublet model \cite{Branco:2011iw}.
  This way the quark and lepton sectors would be clearly independent,
  without any predictions for the CKM matrix.
  In contrast, obtaining successful CKM predictions by assigning
  non-trivial charges in the quark sector constitutes a challenge
  beyond the scope of this paper.
\\[-1.4cm]

\section*{Acknowledgements}

We wish to thank Ernest Ma for many useful discussions and his
insightful comments. RS also wishes to thank the AHEP group at IFIC,
Valencia for its hospitality during his visit, when this work was
initiated.  This work is supported by the Spanish grants FPA2011-2297,
FPA2014-58183-P, Multidark CSD2009-00064, SEV-2014-0398 (MINECO) and
PROMETEOII/2014/084 (Generalitat Valenciana).



\begin{thebibliography}{10}
\providecommand{\url}[1]{\texttt{#1}}
\providecommand{\urlprefix}{URL }
\providecommand{\eprint}[2][]{\url{#2}}

\bibitem{Maltoni:2004ei}
M.~Maltoni, T.~Schwetz, M.~Tortola and J.~Valle, \emph{{Status of global fits
  to neutrino oscillations}},
  \MYhref[journalLinks]{http://dx.doi.org/10.1088/1367-2630/6/1/122}{New
  J.Phys.
  }\MYhref[journalLinks]{http://dx.doi.org/10.1088/1367-2630/6/1/122}{\textbf{6}
  (2004) 122},
  \MYhref[eprintLinks]{http://arxiv.org/abs/hep-ph/0405172}{{\ttfamily
  arXiv:hep-ph/0405172 [hep-ph]}}.

\bibitem{Schechter:1980gr}
J.~Schechter and J.~Valle, \emph{{Neutrino Masses in SU(2) x U(1) Theories}},
  \MYhref[journalLinks]{http://dx.doi.org/10.1103/PhysRevD.22.2227}{Phys.Rev.
  }\MYhref[journalLinks]{http://dx.doi.org/10.1103/PhysRevD.22.2227}{\textbf{D22}
  (1980) 2227}.

\bibitem{Schechter:1980gk}
J.~Schechter and J.~W.~F. Valle, \emph{Neutrino oscillation thought
  experiment}, Phys. Rev. \textbf{D23} (1981) 1666.

\bibitem{Doi:1980ze}
M.~Doi et~al., \emph{{Neutrino Masses and the Double Beta Decay}},
  \MYhref[journalLinks]{http://dx.doi.org/10.1016/0370-2693(81)90746-2}{Phys.
  Lett.
  }\MYhref[journalLinks]{http://dx.doi.org/10.1016/0370-2693(81)90746-2}{\textbf{B103}
  (1981) 219}, [Erratum: Phys. Lett.B113,513(1982)].

\bibitem{avignone:2007fu}
F.~T. Avignone, S.~R. Elliott and J.~Engel, \emph{{Double Beta Decay, Majorana
  Neutrinos, and Neutrino Mass}}, Rev. Mod. Phys. \textbf{80} (2008) 481--516,
  \MYhref[eprintLinks]{http://arxiv.org/abs/0708.1033}{{\ttfamily
  arXiv:0708.1033 [nucl-ex]}}.

\bibitem{Barabash:2004pu}
A.~Barabash, \emph{{Double-beta-decay experiments: Present status and prospects
  for the future}}, Phys.Atom.Nucl. \textbf{67} (2004) 438--452.

\bibitem{Blot:2016cei}
S.~Blot (NEMO-3, SuperNEMO experiments), \emph{{Investigating $\beta\beta$
  decay with the NEMO-3 and SuperNEMO experiments}},
  \MYhref[journalLinks]{http://dx.doi.org/10.1088/1742-6596/718/6/062006}{J.
  Phys. Conf. Ser.
  }\MYhref[journalLinks]{http://dx.doi.org/10.1088/1742-6596/718/6/062006}{\textbf{718}
  (2016) 6 062006}.

\bibitem{Schechter:1981bd}
J.~Schechter and J.~W.~F.~Valle, \emph{{Neutrinoless Double beta Decay in SU(2) x
  U(1) Theories}},
  \MYhref[journalLinks]{http://dx.doi.org/10.1103/PhysRevD.25.2951}{Phys.Rev.
  }\MYhref[journalLinks]{http://dx.doi.org/10.1103/PhysRevD.25.2951}{\textbf{D25}
  (1982) 2951}.

\bibitem{Duerr:2011zd}
M.~Duerr, M.~Lindner and A.~Merle, \emph{{On the Quantitative Impact of the
  Schechter-Valle Theorem}}, JHEP \textbf{1106} (2011) 091,
  \MYhref[eprintLinks]{http://arxiv.org/abs/1105.0901}{{\ttfamily
  arXiv:1105.0901 [hep-ph]}}.

\bibitem{Valle:2015pba}
J.~W.~F.~Valle and J.~C. Romao, \emph{{Neutrinos in high energy and astroparticle
  physics}}, John Wiley \& Sons (2015), ISBN 978-3-527-41197-9.

\bibitem{gell-mann:1980vs}
M.~Gell-Mann, P.~Ramond and R.~Slansky, \emph{Complex spinors and unified
  theories}  (1979), print-80-0576 (CERN).

\bibitem{yanagida:1979}
T.~Yanagida, \emph{Kek lectures}  (KEK lectures, 1979), ed. O. Sawada and A.
  Sugamoto (KEK, 1979).

\bibitem{mohapatra:1980ia}
R.~N. Mohapatra and G.~Senjanovic, \emph{Neutrino mass and spontaneous parity
  nonconservation}, Phys. Rev. Lett. \textbf{44} (1980) 91.

\bibitem{Boucenna:2014zba}
S.~M. Boucenna, S.~Morisi and J.~W.~F.~Valle, \emph{{The low-scale approach to
  neutrino masses}},
  \MYhref[journalLinks]{http://dx.doi.org/10.1155/2014/831598}{Adv.High Energy
  Phys.
  }\MYhref[journalLinks]{http://dx.doi.org/10.1155/2014/831598}{\textbf{2014}
  (2014) 831598},
  \MYhref[eprintLinks]{http://arxiv.org/abs/1404.3751}{{\ttfamily
  arXiv:1404.3751 [hep-ph]}}.


  \bibitem{extra} 
  N.~Arkani-Hamed, S.~Dimopoulos, G.~R.~Dvali and J.~March-Russell,
   \emph{{Neutrino masses from large extra dimensions,}}
  Phys.\ Rev.\ D {\bf 65}, 024032 (2002)
  doi:10.1103/PhysRevD.65.024032; 
  P.~Chen, et al 
  \emph{Warped flavor symmetry predictions for neutrino physics,}
  JHEP {\bf 1601} (2016) 007
  doi:10.1007/JHEP01(2016)007
  [arXiv:1509.06683 [hep-ph]].


\bibitem{Singer:1980sw}
M.~Singer, J.~W.~F.~Valle and J.~Schechter, \emph{{Canonical Neutral Current
  Predictions From the Weak Electromagnetic Gauge Group SU(3) X $u$(1)}},
  \MYhref[journalLinks]{http://dx.doi.org/10.1103/PhysRevD.22.738}{Phys.Rev.
  }\MYhref[journalLinks]{http://dx.doi.org/10.1103/PhysRevD.22.738}{\textbf{D22}
  (1980) 738}.

\bibitem{Addazi:2016xuh}
A.~Addazi, J.~W.~F. Valle and C.~A. Vaquera-Araujo, \emph{{String completion of
  an $\mathrm{SU(3)_c \otimes SU(3)_L \otimes U(1)_X}$ electroweak model}}
  (2016), \MYhref[eprintLinks]{http://arxiv.org/abs/1604.02117}{{\ttfamily
  arXiv:1604.02117 [hep-ph]}}.

\bibitem{Valle:2016kyz}
J.~W.~F. Valle and C.~A. Vaquera-Araujo, \emph{{Dynamical seesaw mechanism for
  Dirac neutrinos}},
  \MYhref[journalLinks]{http://dx.doi.org/10.1016/j.physletb.2016.02.031}{Phys.
  Lett.
  }\MYhref[journalLinks]{http://dx.doi.org/10.1016/j.physletb.2016.02.031}{\textbf{B755}
  (2016) 363--366},
  \MYhref[eprintLinks]{http://arxiv.org/abs/1601.05237}{{\ttfamily
  arXiv:1601.05237 [hep-ph]}}.

\bibitem{Ma:2014qra}
E.~Ma and R.~Srivastava, \emph{{Dirac or inverse seesaw neutrino masses with
  $B-L$ gauge symmetry and $S_3$ flavor symmetry}},
  \MYhref[journalLinks]{http://dx.doi.org/10.1016/j.physletb.2014.12.049}{Phys.
  Lett.
  }\MYhref[journalLinks]{http://dx.doi.org/10.1016/j.physletb.2014.12.049}{\textbf{B741}
  (2015) 217--222},
  \MYhref[eprintLinks]{http://arxiv.org/abs/1411.5042}{{\ttfamily
  arXiv:1411.5042 [hep-ph]}}.

\bibitem{Ma:2015raa}
E.~Ma and R.~Srivastava, \emph{{Dirac or inverse seesaw neutrino masses from
  gauged $B-L$ symmetry}},
  \MYhref[journalLinks]{http://dx.doi.org/10.1142/S0217732315300207}{Mod. Phys.
  Lett.
  }\MYhref[journalLinks]{http://dx.doi.org/10.1142/S0217732315300207}{\textbf{A30}
  (2015) 26 1530020},
  \MYhref[eprintLinks]{http://arxiv.org/abs/1504.00111}{{\ttfamily
  arXiv:1504.00111 [hep-ph]}}.

\bibitem{Ma:2015mjd}
E.~Ma, N.~Pollard, R.~Srivastava and M.~Zakeri, \emph{{Gauge $B-L$ Model with
  Residual $Z_3$ Symmetry}},
  \MYhref[journalLinks]{http://dx.doi.org/10.1016/j.physletb.2015.09.010}{Phys.
  Lett.
  }\MYhref[journalLinks]{http://dx.doi.org/10.1016/j.physletb.2015.09.010}{\textbf{B750}
  (2015) 135--138},
  \MYhref[eprintLinks]{http://arxiv.org/abs/1507.03943}{{\ttfamily
  arXiv:1507.03943 [hep-ph]}}.

\bibitem{Aranda:2013gga}
A.~Aranda et~al., \emph{{Dirac neutrinos from flavor symmetry}},
  \MYhref[journalLinks]{http://dx.doi.org/10.1103/PhysRevD.89.033001}{Phys.
  Rev.
  }\MYhref[journalLinks]{http://dx.doi.org/10.1103/PhysRevD.89.033001}{\textbf{D89}
  (2014) 3 033001},
  \MYhref[eprintLinks]{http://arxiv.org/abs/1307.3553}{{\ttfamily
  arXiv:1307.3553 [hep-ph]}}.

\bibitem{Ma:2006ip}
E.~Ma, \emph{{Neutrino Mass Matrix from Delta(27) Symmetry}},
  \MYhref[journalLinks]{http://dx.doi.org/10.1142/S0217732306021190}{Mod. Phys.
  Lett.
  }\MYhref[journalLinks]{http://dx.doi.org/10.1142/S0217732306021190}{\textbf{A21}
  (2006) 1917--1921},
  \MYhref[eprintLinks]{http://arxiv.org/abs/hep-ph/0607056}{{\ttfamily
  arXiv:hep-ph/0607056 [hep-ph]}}.

\bibitem{Ma:2007wu}
E.~Ma, \emph{{Near tribimaximal neutrino mixing with Delta(27) symmetry}},
  \MYhref[journalLinks]{http://dx.doi.org/10.1016/j.physletb.2007.12.060}{Phys.
  Lett.
  }\MYhref[journalLinks]{http://dx.doi.org/10.1016/j.physletb.2007.12.060}{\textbf{B660}
  (2008) 505--507},
  \MYhref[eprintLinks]{http://arxiv.org/abs/0709.0507}{{\ttfamily
  arXiv:0709.0507 [hep-ph]}}.

\bibitem{Ishimori:2010au}
H.~Ishimori et~al., \emph{{Non-Abelian Discrete Symmetries in Particle
  Physics}}, Prog. Theor. Phys. Suppl. \textbf{183} (2010) 1--163,
  \MYhref[eprintLinks]{http://arxiv.org/abs/1003.3552}{{\ttfamily
  arXiv:1003.3552 [hep-th]}}.

\bibitem{Schechter:1981cv}
J.~Schechter and J.~W.~F. Valle, \emph{{Neutrino Decay and Spontaneous
  Violation of Lepton Number}}, Phys. Rev. \textbf{D25} (1982) 774.

\bibitem{Varzielas:2015aua}
I.~de~Medeiros~Varzielas, \emph{{$\Delta(27)$ family symmetry and neutrino
  mixing}},
  \MYhref[journalLinks]{http://dx.doi.org/10.1007/JHEP08(2015)157}{JHEP
  }\MYhref[journalLinks]{http://dx.doi.org/10.1007/JHEP08(2015)157}{\textbf{08}
  (2015) 157}, \MYhref[eprintLinks]{http://arxiv.org/abs/1507.00338}{{\ttfamily
  arXiv:1507.00338 [hep-ph]}}.

\bibitem{Hernandez:2016eod}
A.~E.~C. Hernández, H.~N. Long and V.~V. Vien, \emph{{A 3-3-1 model with
  right-handed neutrinos based on the $\varDelta \left( 27\right) $ family
  symmetry}},
  \MYhref[journalLinks]{http://dx.doi.org/10.1140/epjc/s10052-016-4074-0}{Eur.
  Phys. J.
  }\MYhref[journalLinks]{http://dx.doi.org/10.1140/epjc/s10052-016-4074-0}{\textbf{C76}
  (2016) 5 242},
  \MYhref[eprintLinks]{http://arxiv.org/abs/1601.05062}{{\ttfamily
  arXiv:1601.05062 [hep-ph]}}.

\bibitem{Vien:2016tmh}
V.~V. Vien, A.~E.~C. Hernandez and H.~N. Long, \emph{{The $\Delta(27)$ flavor
  3-3-1 model with neutral leptons}}  (2016),
  \MYhref[eprintLinks]{http://arxiv.org/abs/1601.03300}{{\ttfamily
  arXiv:1601.03300 [hep-ph]}}.

\bibitem{Forero:2014bxa}
D.~Forero, M.~Tortola and J.~W.~F.~Valle, \emph{{Neutrino oscillations refitted}},
  \MYhref[journalLinks]{http://dx.doi.org/10.1103/PhysRevD.90.093006}{Phys.Rev.
  }\MYhref[journalLinks]{http://dx.doi.org/10.1103/PhysRevD.90.093006}{\textbf{D90}
  (2014) 9 093006},
  \MYhref[eprintLinks]{http://arxiv.org/abs/1405.7540}{{\ttfamily
  arXiv:1405.7540 [hep-ph]}}.

\bibitem{Ade:2015xua}
P.~A.~R. Ade et~al. (Planck), \emph{{Planck 2015 results. XIII. Cosmological
  parameters}}  (2015),
  \MYhref[eprintLinks]{http://arxiv.org/abs/1502.01589}{{\ttfamily
  arXiv:1502.01589 [astro-ph.CO]}}.

\bibitem{Rodejohann:2011vc}
W.~Rodejohann and J.~W.~F. Valle, \emph{{Symmetrical Parametrizations of the
  Lepton Mixing Matrix}}, Phys.Rev. \textbf{D84} (2011) 073011,
  \MYhref[eprintLinks]{http://arxiv.org/abs/1108.3484}{{\ttfamily
  arXiv:1108.3484 [hep-ph]}}.

\bibitem{Chulia:2016ngi}
S.~C. Chuliá, E.~Ma, R.~Srivastava and J.~W.~F. Valle, \emph{{Dirac Neutrinos
  and Dark Matter Stability from Lepton Quarticity}}  (2016),
  \MYhref[eprintLinks]{http://arxiv.org/abs/1606.04543}{{\ttfamily
  arXiv:1606.04543 [hep-ph]}}.

\bibitem{Akerib:2016vxi}
D.~S.~Akerib et~al. (LUX Collaboration), \emph{{Results from a search for dark matter in LUX with 332 live days of exposure}} (2016),
    \MYhref[eprintLinks]{http://arxiv.org/abs/1608.07648}{{\ttfamily
  arXiv:1608.07648 [astro-ph.CO]}}.
 
 
\bibitem{Tan:2016zwf}
A.~Tan et~al. (PandaX-II Collaboration), \emph{{Dark Matter Results from First 98.7-day Data of PandaX-II Experiment}} (2016),
  \MYhref[eprintLinks]{http://arxiv.org/abs/1607.07400}{{\ttfamily
  arXiv:1607.07400 [hep-ex]}}.
  
\bibitem{Aad:2015txa}
G.~Aad et~al. (ATLAS), \emph{{Search for invisible decays of a Higgs boson
  using vector-boson fusion in $pp$ collisions at $\sqrt{s}=8$ TeV with the
  ATLAS detector}},
  \MYhref[journalLinks]{http://dx.doi.org/10.1007/JHEP01(2016)172}{JHEP
  }\MYhref[journalLinks]{http://dx.doi.org/10.1007/JHEP01(2016)172}{\textbf{01}
  (2016) 172}, \MYhref[eprintLinks]{http://arxiv.org/abs/1508.07869}{{\ttfamily
  arXiv:1508.07869 [hep-ex]}}.

\bibitem{Boucenna:2011tj}
M.~Boucenna et~al., \emph{{Phenomenology of Dark Matter from $A_4$ Flavor
  Symmetry}}, JHEP \textbf{1105} (2011) 037,
  \MYhref[eprintLinks]{http://arxiv.org/abs/1101.2874}{{\ttfamily
  arXiv:1101.2874 [hep-ph]}}.

\bibitem{Khachatryan:2014jba}
V.~Khachatryan et~al. (CMS), \emph{{Precise determination of the mass of the
  Higgs boson and tests of compatibility of its couplings with the standard
  model predictions using proton collisions at 7 and 8 $\,\text {TeV}$}},
  \MYhref[journalLinks]{http://dx.doi.org/10.1140/epjc/s10052-015-3351-7}{Eur.
  Phys. J.
  }\MYhref[journalLinks]{http://dx.doi.org/10.1140/epjc/s10052-015-3351-7}{\textbf{C75}
  (2015) 5 212},
  \MYhref[eprintLinks]{http://arxiv.org/abs/1412.8662}{{\ttfamily
  arXiv:1412.8662 [hep-ex]}}.

\bibitem{Cline:2013gha}
J.~M. Cline, K.~Kainulainen, P.~Scott and C.~Weniger, \emph{{Update on scalar
  singlet dark matter}},
  \MYhref[journalLinks]{http://dx.doi.org/10.1103/PhysRevD.92.039906,
  10.1103/PhysRevD.88.055025}{Phys. Rev.
  }\MYhref[journalLinks]{http://dx.doi.org/10.1103/PhysRevD.92.039906,
  10.1103/PhysRevD.88.055025}{\textbf{D88} (2013) 055025}, [Erratum: Phys.
  Rev.D92,no.3,039906(2015)],
  \MYhref[eprintLinks]{http://arxiv.org/abs/1306.4710}{{\ttfamily
  arXiv:1306.4710 [hep-ph]}}.

\bibitem{Feng:2014vea}
L.~Feng, S.~Profumo and L.~Ubaldi, \emph{{Closing in on singlet scalar dark
  matter: LUX, invisible Higgs decays and gamma-ray lines}},
  \MYhref[journalLinks]{http://dx.doi.org/10.1007/JHEP03(2015)045}{JHEP
  }\MYhref[journalLinks]{http://dx.doi.org/10.1007/JHEP03(2015)045}{\textbf{03}
  (2015) 045}, \MYhref[eprintLinks]{http://arxiv.org/abs/1412.1105}{{\ttfamily
  arXiv:1412.1105 [hep-ph]}}.

\bibitem{Chen:2012jg}
M.-C. Chen, M.~Ratz, C.~Staudt and P.~K.~S. Vaudrevange, \emph{{The mu Term and
  Neutrino Masses}},
  \MYhref[journalLinks]{http://dx.doi.org/10.1016/j.nuclphysb.2012.08.018}{Nucl.
  Phys.
  }\MYhref[journalLinks]{http://dx.doi.org/10.1016/j.nuclphysb.2012.08.018}{\textbf{B866}
  (2013) 157--176},
  \MYhref[eprintLinks]{http://arxiv.org/abs/1206.5375}{{\ttfamily
  arXiv:1206.5375 [hep-ph]}}.

\bibitem{Heeck:2013rpa}
J.~Heeck and W.~Rodejohann, \emph{{Neutrinoless Quadruple Beta Decay}},
  \MYhref[journalLinks]{http://dx.doi.org/10.1209/0295-5075/103/32001}{Europhys.
  Lett.
  }\MYhref[journalLinks]{http://dx.doi.org/10.1209/0295-5075/103/32001}{\textbf{103}
  (2013) 32001},
  \MYhref[eprintLinks]{http://arxiv.org/abs/1306.0580}{{\ttfamily
  arXiv:1306.0580 [hep-ph]}}.

\bibitem{Heeck:2013vha}
J.~Heeck, \emph{{Leptogenesis with Lepton-Number-Violating Dirac Neutrinos}},
  \MYhref[journalLinks]{http://dx.doi.org/10.1103/PhysRevD.88.076004}{Phys.
  Rev.
  }\MYhref[journalLinks]{http://dx.doi.org/10.1103/PhysRevD.88.076004}{\textbf{D88}
  (2013) 076004},
  \MYhref[eprintLinks]{http://arxiv.org/abs/1307.2241}{{\ttfamily
  arXiv:1307.2241 [hep-ph]}}.

\bibitem{Bonilla:2016sgx}
C.~Bonilla, M.~Nebot, R.~Srivastava and J.~W.~F. Valle, \emph{{Flavor physics
  scenario for the 750 GeV diphoton anomaly}},
  \MYhref[journalLinks]{http://dx.doi.org/10.1103/PhysRevD.93.073009}{Phys.
  Rev.
  }\MYhref[journalLinks]{http://dx.doi.org/10.1103/PhysRevD.93.073009}{\textbf{D93}
  (2016) 7 073009},
  \MYhref[eprintLinks]{http://arxiv.org/abs/1602.08092}{{\ttfamily
  arXiv:1602.08092 [hep-ph]}}.

\bibitem{Modak:2016ung}
T.~Modak, S.~Sadhukhan and R.~Srivastava, \emph{{750 GeV diphoton excess from
  gauged $B-L$ symmetry}},
  \MYhref[journalLinks]{http://dx.doi.org/10.1016/j.physletb.2016.03.021}{Phys.
  Lett.
  }\MYhref[journalLinks]{http://dx.doi.org/10.1016/j.physletb.2016.03.021}{\textbf{B756}
  (2016) 405--412},
  \MYhref[eprintLinks]{http://arxiv.org/abs/1601.00836}{{\ttfamily
  arXiv:1601.00836 [hep-ph]}}.
  
  
\bibitem{Branco:2011iw}
G.~C.~Branco, P.~M.~Ferreira, L.~Lavoura, M.~N.~Rebelo, Marc Sher, Joao P.~Silva, \emph{{Theory and phenomenology of two-Higgs-doublet models}},
  \MYhref[journalLinks]{http://dx.doi.org/10.1016/j.physrep.2012.02.002}{Phys. Rept.
  }\MYhref[journalLinks]{http://dx.doi.org/10.1016/j.physrep.2012.02.002}{\textbf{516}
  (2012) 1--102},
  \MYhref[eprintLinks]{https://arxiv.org/abs/1106.0034}{{\ttfamily
  arXiv:1106.0034 [hep-ph]}}.
  

\end{thebibliography}
\end{document}